\documentclass[preprint,prd,aps,showpacs,showkeys,nofootinbib]{revtex4}
\usepackage{graphicx}
\begin{document}



\title{The study of the $Z\rightarrow l_i^{\pm}l_j^{\mp}$ processes in the BLMSSM}

\author{Xing-Xing Dong$^{1,}$\footnote{dxx-0304@163.com},
Shu-Min Zhao$^{1,}$\footnote{zhaosm@hbu.edu.cn},
 Xi-Jie Zhan$^{1}$,\\
Zhong-Jun Yang$^{1}$,
Hai-Bin Zhang$^{1}$,
Tai-Fu Feng$^{1}$}

\affiliation{$^1$ Department of Physics, Hebei University, Baoding, 071002, China}

\begin{abstract}
In a supersymmetric extension of the Standard Model(SM) where baryon and lepton numbers are local gauge symmetries (BLMSSM), we investigate the charged lepton flavor violating (CLFV) processes $Z\rightarrow l_i^{\pm}l_j^{\mp}$ after introducing the new gauginos and the right-handed neutrinos. In this model, the branching ratios of $Z\rightarrow l_i^{\pm}l_j^{\mp}$ are around ($10^{-8}\sim10^{-10}$), which approach the present experimental upper bounds. We hope that the branching ratios for these CLFV processes can be detected in the near future.
\end{abstract}


\maketitle

\section{Introduction\label{sec1}}
Neutrinos have tiny masses and mix with each other, which can be proved by the neutrino oscillation experiments\cite{neutrino1,neutrino2,neutrino3,neutrino4}. It shows that lepton flavor symmetry is not conserved in neutrino sector. A new particle around 125 GeV is detected by the LHC\cite{LHC1,LHC2,LHC3}, whose properties are close to the Higgs boson. Then the SM has achieved great success. However, due to the GIM mechanism, the expected rates for the charged lepton flavor violating (CFLV) processes\cite{CLFV1,CLFV2} are very tiny in the SM with massive neutrinos. For example, $Br(Z\rightarrow e\mu)\sim Br(Z\rightarrow e\tau)\sim10^{-54}$ and $Br(Z\rightarrow\mu\tau)\sim10^{-60}$\cite{BrZ1,BrZ2,BrZ4,BrZ5,BrZ6}, they are much smaller than the experimental upper bounds. The CLFV is forbidden in the SM. In Table 1, we show the present experimental limits and future sensitivities for some CLFV processes. In the Ref.\cite{CEPC}, the authors consider that the future sensitivities for the CLFV processes may be reached $10^{-11}$; At a Future Circular $e^+e^-$ Collider (such as FCC-ee(TLEP))\cite{TLEP1,TLEP2}, it is estimated that the sensitivities can be improved up to $10^{-13}$. Thus, any signal of CLFV would be a hint of new physics, and the study of CLFV processes is an effective approach to explore new physics beyond SM.
\begin{table}[h]
\caption{ \label{tab1}  Present experimental limits and future sensitivities for the CLFV processes $Z\rightarrow l_i^{\pm}l_j^{\mp}$.}
\footnotesize
\begin{tabular*}{150mm}{@{\extracolsep{\fill}}ccc}
\toprule  CLFV process& Present limit  & Future sensitivity(TESLA) \\
\hline
$Z\rightarrow e\mu$ \hphantom{00} & \hphantom{0}$<7.5\times10^{-7}$ \cite{PDG1,ATLAS,CMS}& \hphantom{0}$\sim2.0\times10^{-9}$\cite{TESLA} \\
$Z\rightarrow e\tau$ \hphantom{00} & \hphantom{0}$<9.8\times10^{-6}$\cite{PDG1,PDG2,PDG3} & \hphantom{0}$\sim(1.3-6.5)\times10^{-8}$\cite{TESLA} \\
$Z\rightarrow \mu\tau$ \hphantom{00} & \hphantom{0}$<1.2\times10^{-5}$\cite{PDG1,PDG2,PDG4} &\hphantom{0} $\sim(0.44-2.2)\times10^{-8}$\cite{TESLA} \\
\hline
\end{tabular*}%
\end{table}

In simple SM extension, CLFV processes are restricted strongly by the tiny neutrino masses. As an appealing supersymmetric extension of SM, the minimal supersymmetric standard model (MSSM)\cite{MSSM1,MSSM2,MSSM3,MSSM4} with R-party\cite{MSSM3} conservation has drawn physicists' attention for a long time. However, the left-handed light neutrinos remain massless, and it can not explain the discovery of neutrino oscillations. Therefore, physicists do more research on the light neutrino masses and mixings with MSSM extension\cite{MSSMe1,MSSMe2,MSSMe3,MSSMe4,SSM1,SSM2}. As a supersymmetric extension of the MSSM with local gauged baryon $(B)$ and lepton $(L)$ numbers, BLMSSM is introduced\cite{BLMSSM1,BLMSSM2,BLMSSM3,BLMSSM4}.
In the BLMSSM, the local gauged $B$ must be broken in order to account for the asymmetry of matter-antimatter in the universe. Right-handed neutrinos are introduced to explain the data from neutrino oscillation experiments, hence lepton number is also expected to be broken\cite{BLMSSM2}. In Refs.\cite{BLMSSM2,BL}, baryon number and lepton number are local gauged and spontaneously broken at the TeV scale in the BLMSSM.

In this work, we continue to analyze the CLFV processes $Z\rightarrow l_i^{\pm}l_j^{\mp} (Z\rightarrow e\mu, Z\rightarrow e\tau, Z\rightarrow \mu\tau )$ within the BLMSSM. Compared with the MSSM, the neutrino masses in the BLMSSM are not zero. Three heavy neutrinos and three new scalar neutrinos are introduced in this model. And new particle lepton neutralino $\chi_L^0$ is also introduced. These new sources enlarge the CLFV processes via loop contributions. Therefore, the expected experimental results for the CLFV processes may be obtained in the near future.

This work is organized as follows: In Sec.2, we summarize the BLMSSM briefly, including its superpotential, the general soft SUSY-breaking terms, needed mass matrices and couplings. Section 3 is devoted to the decay widths of the CLFV processes $Z\rightarrow l_i^{\pm}l_j^{\mp}$. In Sec.4, we give out the corresponding parameters and numerical analysis. The discussion and conclusion are described in Section 5. Appendix A is devoted to described the concrete forms of coupling coefficients in Fig.1.

\section{BLMSSM\label{sec2}}
The local gauge group of BLMSSM $SU(3)_C\otimes{SU(2)_L}\otimes{U(1)_Y}\otimes{U(1)_B}\otimes{U(1)_L}$\cite{BLMSSM1,SU1,SU2} enlarges the SM. In the BLMSSM, the new quarks superfields $\hat{Q}_4\sim(3,2,1/6,B_4,0)$, $\hat{U}_4^c\sim (\overline{3},1,-2/3,-B_4,0)$, $\hat{D}_4^c\sim(\overline{3},1,1/3,-B_4,0)$, $\hat{Q}_5^c\sim(\overline{3},2,-1/6,-(1+B_4),0)$, $\hat{U}_5\sim (3,1,2/3,1+B_4,0)$ and  $\hat{D}_5\sim(3,1-1/3,1+B_4,0)$ are introduced to cancel B anomaly. To break baryon number spontaneously, the model introduces Higgs superfields $\hat{\Phi}_B\sim(1,1,0,1,0)$ and $\hat{\varphi}_B\sim(1,1,0,-1,0)$. The new leptons superfields $\hat{L}_4\sim(1,2,-1/2,0,L_4)$, $\hat{E}_4^c\sim(1,1,1,0,-L_4)$, $\hat{N}_4^c\sim(1,1,0,0,-L_4)$, $\hat{L}_5^c\sim(1,2,1/2,0,-(3+L_4))$, $\hat{E}_5\sim(1,1-1,0,3+L_4)$ and $\hat{N}_5\sim(1,1,0,0,3+L_4)$ are introduced to cancel L anomaly. The exotic Higgs superfields $\hat{\Phi}_L\sim(1,1,0,0,-2)$ and $\hat{\varphi}_L\sim(1,1,0,0,2)$ can break lepton number spontaneously. Here $B_4$ and $L_4$ stand for baryon and lepton numbers for a given field respectively. In our numerical calculation, we use $B_4=3/2$ and $L_4=3/2$.
 The exotic Higgs superfields $\hat{\Phi}_B$, $\hat{\varphi}_B$ and $\hat{\Phi}_L$, $\hat{\varphi}_L$ acquire nonzero vacuum expectation values (VEVs), then the exotic quarks and exotic leptons obtain masses. The model also includes the superfields
$\hat{X}\sim(1,1,0,2/3+B_4,0)$ and $\hat{X}^{'}\sim(1,1,0,-(2/3+B_4),0)$ to make exotic quarks unstable. Furthermore, with $\hat{X}$ and $\hat{X}^{'}$ mixing together, the lightest mass eigenstate can be a dark matter candidate.

The superpotential of the BLMSSM is shown as follows\cite{superpotential}
\begin{eqnarray}
&&{\cal W}_{BLMSSM}={\cal W}_{MSSM}+{\cal W}_{B}+{\cal W}_{L}+{\cal W}_{X}\;,
\end{eqnarray}
with ${\cal W}_{MSSM}$ representing the superpotential of the MSSM. The concrete forms of ${\cal W}_{B}$, ${\cal W}_{L}$ and ${\cal W}_{X}$ can be obtained in Ref.\cite{superpotential}.

In the BLMSSM, the soft breaking terms $\mathcal{L}_{soft}$ are generally given by\cite{BLMSSM1,BLMSSM2,superpotential}, and only the leptonic terms contribute to our study
\begin{eqnarray}
&&{\cal L}_{{soft}}=-(m_{{\tilde{N}^c}}^2)_{{IJ}}\tilde{N}_I^{c*}\tilde{N}_J^c-m_{{\Phi_{L}}}^2\Phi_{L}^*\Phi_{L}
-m_{{\varphi_{L}}}^2\varphi_{L}^*\varphi_{L}
-\Big(m_{L}\lambda_{L}\lambda_{L}+h.c.\Big)\nonumber\\&&\hspace{1.1cm}+A_{N}Y_{\nu}\tilde{L}H_{u}\tilde{N}^c
+A_{{N^c}}\lambda_{{N^c}}\tilde{N}^c\tilde{N}^c\varphi_{L}
+B_{L}\mu_{L}\Phi_{L}\varphi_{L}+h.c.\Big\}.
\end{eqnarray}
Here $\lambda_L$ represents gaugino of $U(1)_L$. The $SU(2)_L$ doublets $H_{u}$ and $H_{d}$ obtain the nonzero VEVs $\upsilon_{u}$ and $\upsilon_{d}$,
\begin{eqnarray}
&&H_u=\left({\begin{array}{*{20}{c}}
H_u^+  \\
\frac{1}{\sqrt 2}(\upsilon_u+H_u^0+iP_u^0)  \\
\end{array}}
\right), \nonumber\\
&&H_d=\left({\begin{array}{*{20}{c}}
\frac{1}{\sqrt 2}(\upsilon_d+H_d^0+iP_d^0)  \\
H_d^-  \\
\end{array}}
\right),
\end{eqnarray}
The $SU(2)_L$ singlets $\Phi_{L}$ and $\varphi_{L}$ acquire the nonzero VEVs $\upsilon_{L}$ and $\overline{\upsilon}_{L}$,
\begin{eqnarray}
&&\Phi_L=\frac{1}{\sqrt{2}}(\upsilon_L+\Phi_L^0+iP_L^0),\nonumber\\
&&\varphi_L=\frac{1}{\sqrt{2}}
(\overline{\upsilon}_L+\varphi_L^0+i\overline{P}_L^0).
\end{eqnarray}

In the BLMSSM, the mass matrices of lepton neutralinos, neutrinos, sleptons and sneutrinos are introduced as follows:

In the base $(i\lambda_L, \psi_{\Phi_L}, \psi_{\varphi_L})$\cite{BLMSSM4,matrix,base1}, the mixing mass matrix of lepton neutralinos is obtained.
\begin{equation}
M_{LN}=\left(\begin{array}{ccc}
  2M_L &2v_Lg_L &-2\bar{v}_Lg_L\\
   2v_Lg_L & 0 &-\mu_L\\-2\bar{v}_Lg_L&-\mu_L &0
    \end{array}\right).
   \end{equation}
Then the three lepton neutralino masses are deduced due to diagonalize the mass matrix $M_{LN}$ by $Z_{N_L}$

After symmetry breaking, the mass matrix of neutrinos is deduced in the basis $(\nu, N^c)$\cite{basis1,basis2}
\begin{eqnarray}
\left(\begin{array}{cc}
  0&\frac{v_u}{\sqrt{2}}(Y_{\nu})_{IJ} \\
   \frac{v_u}{\sqrt{2}}(Y^{T}_{\nu})_{IJ}  & \frac{\bar{v}_L}{\sqrt{2}}(\lambda_{N^c})_{IJ}
    \end{array}\right).
\end{eqnarray}
 Then diagonalizing the neutrino mass matrix by the unitary matrix $U_{\nu}$, we can get six mass eigenstates of neutrinos, which include three light eigenstates and three heavy eigenstates.

In the BLMSSM, the slepton mass squared matrix deduced from Eqs.(1),(2) reads as
\begin{eqnarray}
&&\left(\begin{array}{cc}
  (\mathcal{M}^2_L)_{LL}&(\mathcal{M}^2_L)_{LR} \\
   (\mathcal{M}^2_L)_{LR}^{\dag} & (\mathcal{M}^2_L)_{RR}
    \end{array}\right),
\end{eqnarray}
where,
\begin{eqnarray}
 &&(\mathcal{M}^2_L)_{LL}=\frac{(g_1^2-g_2^2)(v_d^2-v_u^2)}{8}\delta_{IJ} +g_L^2(\bar{v}_L^2-v_L^2)\delta_{IJ}
+m_{l^I}^2\delta_{IJ}+(m^2_{\tilde{L}})_{IJ},\nonumber\\&&
 (\mathcal{M}^2_L)_{LR}=\frac{\mu^*v_u}{\sqrt{2}}(Y_l)_{IJ}-\frac{v_u}{\sqrt{2}}(A'_l)_{IJ}+\frac{v_d}{\sqrt{2}}(A_l)_{IJ},
 \nonumber\\&& (\mathcal{M}^2_L)_{RR}=\frac{g_1^2(v_u^2-v_d^2)}{4}\delta_{IJ}-g_L^2(\bar{v}_L^2-v_L^2)\delta_{IJ}
+m_{l^I}^2\delta_{IJ}+(m^2_{\tilde{R}})_{IJ}.
\end{eqnarray}
Through the matrix $Z_{\tilde{L}}$, the mass matrix can be diagonalized.

From the contributions of Eqs.(1),(2), we also deduce the mass squared matrix of sneutrino ${\cal M}_{\tilde{n}}$ with $\tilde{n}^T=(\tilde{\nu},\tilde{N}^c)$
\begin{eqnarray}
&&\left(\begin{array}{cc}
  {\cal M}^2_{\tilde{n}}(\tilde{\nu}_{I}^*\tilde{\nu}_{J})&{\cal M}^2_{\tilde{n}}(\tilde{\nu}_I\tilde{N}_J^c) \\
   ({\cal M}^2_{\tilde{n}}(\tilde{\nu}_I\tilde{N}_J^c))^{\dag} & {\cal M}^2_{\tilde{n}}(\tilde{N}_I^{c*}\tilde{N}_J^c)
    \end{array}\right),
\end{eqnarray}
where,
\begin{eqnarray}
  && {\cal M}^2_{\tilde{n}}(\tilde{\nu}_{I}^*\tilde{\nu}_{J})=\frac{g_1^2+g_2^2}{8}(v_d^2-v_u^2)\delta_{IJ}+g_L^2(\overline{v}^2_L-v^2_L)\delta_{IJ}
 +\frac{v_u^2}{2}(Y^\dag_{\nu}Y_\nu)_{IJ}+(m^2_{\tilde{L}})_{IJ},\nonumber\\&&
   {\cal M}^2_{\tilde{n}}(\tilde{\nu}_I\tilde{N}_J^c)=\mu^*\frac{v_d}{\sqrt{2}}(Y_{\nu})_{IJ}-v_u\overline{v}_L(Y_{\nu}^\dag\lambda_{N^c})_{IJ}
 +\frac{v_u}{\sqrt{2}}(A_{N})_{IJ}(Y_\nu)_{IJ},\nonumber\\&&
   {\cal M}^2_{\tilde{n}}(\tilde{N}_I^{c*}\tilde{N}_J^c)=-g_L^2(\overline{v}^2_L-v^2_L)\delta_{IJ}
   +\frac{v_u^2}{2}(Y^\dag_{\nu}Y_\nu)_{IJ}
+2\overline{v}^2_L(\lambda_{N^c}^\dag\lambda_{N^c})_{IJ}
\nonumber\\&&\hspace{2.7cm}+\mu_L\frac{v_L}{\sqrt{2}}(\lambda_{N^c})_{IJ}
+(m^2_{\tilde{N}^c})_{IJ}-\frac{\overline{v}_L}{\sqrt{2}}(A_{N^c})_{IJ}(\lambda_{N^c})_{IJ}.
   \end{eqnarray}
Then the sneutrino masses can be obtained by formula
$Z_{\nu^{IJ}}^{\dag} {\cal M}_{\tilde{n}}^2Z_{\nu^{IJ}}=diag(m_{\tilde{\nu}_1^1}^2, m_{\tilde{\nu}_1^2}^2, m_{\tilde{\nu}_1^3}^2, m_{\tilde{\nu}_2^1}^2, m_{\tilde{\nu}_2^2}^2, m_{\tilde{\nu}_2^3}^2)$.

In the BLMSSM, we deduce the corrections for the couplings existed in the MSSM due to superfields $\tilde{N}^c$. The corresponding couplings for W-lepton-neutrino,  Z-neutrino-neutrino, charged Higgs-lepton-neutrino, Z-sneutrino-sneutrino and chargino-lepton-sneutrino are introduced in Ref.\cite{BLMSSM4}.

From the interactions of gauge and matter multiplets
$ig\sqrt{2}T^a_{ij}(\lambda^a\psi_jA_i^*-\bar{\lambda}^a\bar{\psi}_iA_j)$,
the lepton-slepton-lepton neutralino coupling is deduced here
\begin{eqnarray}
&&\mathcal{L}_{l\chi_L^0\tilde{L}}=\sqrt{2}g_L\bar{\chi}_{L_j}^0\Big(Z_{N_L}^{1j}Z_{L}^{Ii}P_L
-Z_{N_L}^{1j*}Z_{L}^{(I+3)i}P_R\Big)l^I\tilde{L}_i^++h.c.
\end{eqnarray}

\section{The CLFV decays $Z\rightarrow l_i^{\pm}l_j^{\mp}$\label{sec3}}
In the BLMSSM, we study the CLFV processes $Z\rightarrow l_i^{\pm}l_j^{\mp}$. The corresponding Feynman diagrams can be depicted by Fig.1,
and the corresponding effective amplitudes can be written as \cite{BrZ6,SSM3,amplitude3}
\begin{eqnarray}
{\cal M}_{\mu}={\bar l}_i\gamma_{\mu}(F_LP_L+F_RP_R)l_j,
\end{eqnarray}
with
\begin{eqnarray}
F_{L,R}=F_{L,R}(S)+F_{L,R}(W),
\end{eqnarray}
where $l_{i,j}$ represent the wave functions of the external leptons. The coefficients $F_{L,R}$ can be obtained from the amplitudes of the Feynman diagrams. $F_{L,R}(S)$ correspond to
\begin{figure}[ht]
\centering
\includegraphics[width=12cm]{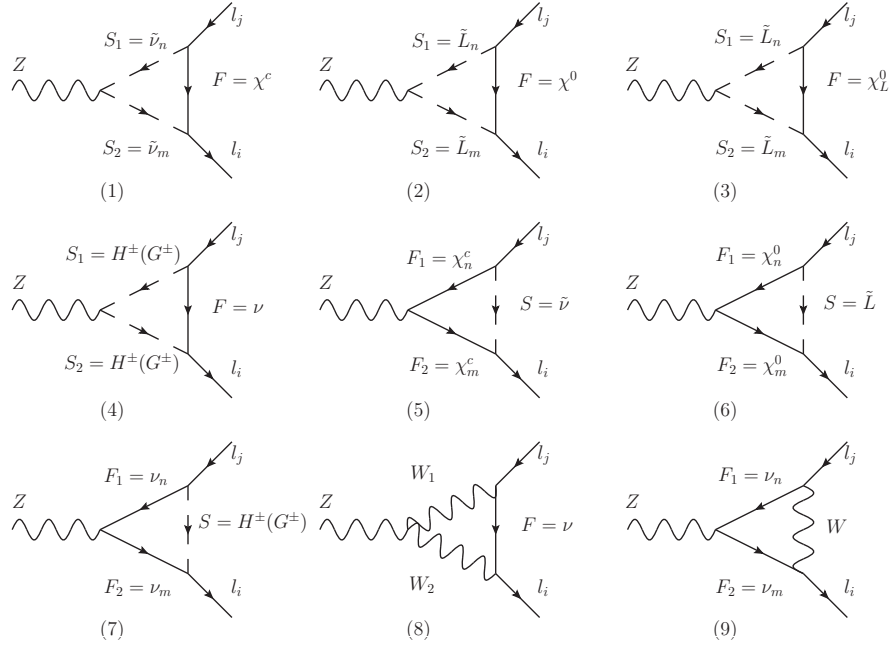}
\caption{Feynman diagrams for the $Z\rightarrow l_i^{\pm}l_j^{\mp}$ processes in the BLMSSM: F represents Dirac(Majorana) fermion particle, S represents scalar boson particle and W represents W boson particle.} \label{fig1}
\end{figure}
Fig.1(1)$\sim$Fig.1(7), and stand for the contributions from chargino-sneutrino, neutralino-slepton, neutrino-charged Higgs and lepton neutralino--slepton; $F_{L,R}(W)$ correspond to Fig.1(8) and Fig.1(9), and stand for the contributions from W-neutrino due to three light neutrinos and three heavy neutrinos mixing together. We formulate these coefficients as follows
\begin{eqnarray}
&&F_L(S)=\frac{i}{2}\sum_{F=\chi^c,\chi^0,\nu}\sum_{S=\tilde{\nu},\tilde{L},H^{\pm}(G^{\pm})}
\big[\frac{2m_{_{F_1}}m_{_{F_2}}}{m_{N_p}^2}H_R^{SF_2\bar{l}_i}H_L^{ZF_1\bar{F}_2}H_L^{S^*l_j\bar{F}_1}G_1(x_S,x_{F_1},x_{F_2})
\nonumber\\&&\hspace{1.7cm}-H_R^{SF_2\bar{l}_i}H_R^{ZF_1\bar{F}_2}H_L^{S^*l_j\bar{F}_1}G_2(x_S,x_{F_1},x_{F_2})
+H_R^{S_2F\bar{l}_i}H^{ZS_1S_2^*}H_L^{S_1^*l_j\bar{F}}G_2(x_F,x_{S_1},x_{S_2})\big]
\nonumber\\&&\hspace{1.7cm}+\frac{i}{2}\sum_{F=\chi_L^0}\sum_{S=\tilde{L}}
[H_R^{S_2F\bar{l}_i}H^{ZS_1S_2^*}H_L^{S_1^*l_j\bar{F}}G_2(x_F,x_{S_1},x_{S_2})],
\nonumber\\&&F_R(S)=F_L(S)|_{L{\leftrightarrow}R};
\nonumber\\&&F_L(W)=i\sum_{F=\nu}\sum_{W=W_{\mu}}[3H_L^{W_2F\bar{l}_i}H^{ZW_1W_2^*}H_L^{W_1^*l_j\bar{F}}G_2(x_F,x_{W_1},x_{W_2})
\nonumber\\&&\hspace{1.7cm}-H_L^{WF_2\bar{l}_i}H_L^{ZF_1\bar{F}_2}H_L^{\bar{F}_1l_jW^*}G_2(x_W,x_{F_1},x_{F_2})],
\nonumber\\&&F_R(W)=0.
\end{eqnarray}
Here, $H_{L,R}^{SF_2\bar{l}_i}...$ represent the corresponding coupling coefficients of the left (right)-hand parts in the Lagrangian and the concrete expressions can be found in Appendix. $x_i=\frac{m^2}{m_{N_p}^2}$ with $m$ representing the mass of the corresponding particle, $m_{N_p}$ representing energy scale of the new physics to make the amplitudes dimensionless. The one-loop functions $G_i(x_1,x_2,x_3), i=1,2 $ are given by
\begin{eqnarray}
&&\hspace{-0.7cm}G_1(x_1,x_2,x_3)=\frac{1}{16{\pi}^2}[\frac{x_1\ln{x_1}}{(x_1-x_2)(x_1-x_3)}+\frac{x_2\ln{x_2}}{(x_2-x_1)(x_2-x_3)}
+\frac{x_3\ln{x_3}}{(x_3-x_1)(x_3-x_2)}],
\nonumber\\
&&\hspace{-0.7cm}G_2(x_1,x_2,x_3)=\frac{1}{16{\pi}^2}[\frac{x_1^2\ln{x_1}}{(x_1-x_2)(x_1-x_3)}
+\frac{x_2^2\ln{x_2}}{(x_2-x_1)(x_2-x_3)}+\frac{x_3^2\ln{x_3}}{(x_3-x_1)(x_3-x_2)}].
\end{eqnarray}

Then, the branching ratios of $Z\rightarrow l_i^{\pm}l_j^{\mp}$ can be summarized as
\begin{eqnarray}
Br\left(Z\rightarrow l_i^{\pm}l_j^{\mp}\right)=\frac{1}{12\pi}\frac{m_Z}{\Gamma_Z}\left(|F_L|^2+|F_R|^2\right)
=\frac{1}{12\pi}\frac{m_Z}{\Gamma_Z}\left(|F_L(S)+F_L(W)|^2
+|F_R(S)|^2\right),
\end{eqnarray}
where $\Gamma_Z$ represents the total decay width of Z-boson and we use $\Gamma_Z\simeq2.4952$ GeV\cite{PDG1}.

\section{Numerical Results for the CLFV processes $Z\rightarrow l_i^{\pm}l_j^{\mp}$\label{sec4}}
In this section, we study the numerical results, and consider the experiment constraints from the light neutral Higgs mass $m_{_{h^0}}\simeq125\;{\rm GeV}$ \cite{PDG1,LHC1,LHC2,LHC3} and the neutrino experiment data\cite{PDG1}
\begin{eqnarray}
&&\sin^2\theta_{13}= (2.19\pm 0.12)\times10^{-2},\sin^2\theta_{12} =0.304\pm0.014, \sin^2\theta_{23}=0.51\pm0.05,
\nonumber\\&&\Delta m_{\odot}^2 =(7.53\pm 0.18)\times 10^{-5} {\rm eV}^2,
|\Delta m_{A}^2| =(2.44\pm0.06)\times 10^{-3} {\rm eV}^2.
\end{eqnarray}
In our previous works, the neutron EDM, muon MDM and lepton EDM are studied\cite{base1,basis1,matrix}, whose constraints are taken into account here. In the Refs.\cite{PDG1,BLMSSM4}, $Br(\mu \to e+\gamma) < 5.7\times10^{-13}$ and $Br(\mu \to 3e)<1.0\times10^{-12}$ are strict limits for our parameter space. Furthermore, the ratios for $h\rightarrow\gamma\gamma$, $h\rightarrow ZZ^*$ and $h\rightarrow WW^*$ are around $1.16\pm0.18$, $1.29_{-0.23}^{+0.26}$ and $1.08_{-0.16}^{+0.18}$ respectively\cite{PDG1}, which are also considered in our parameter space. In this work, the used parameters are given out\cite{PDG1,matrix,basis1}:
\begin{eqnarray}
&&m_e=0.51\times10^{-3}{\rm GeV},m_Z=91.1876{\rm GeV},
m_{\mu}=0.105{\rm GeV},\nonumber\\&&m_{\tau}=1.777{\rm GeV},
m_W=80.385{\rm GeV},
\alpha(m_Z)=1/128,s_W^2(m_Z)=0.23,
\nonumber\\&&(Y_{\nu})_{11}=1.3031*10^{-6},(Y_{\nu})_{12}=9.0884*10^{-8},
(Y_{\nu})_{13}=6.9408*10^{-8},\nonumber\\&&(Y_{\nu})_{22}=1.6002*10^{-6},
(Y_{\nu})_{23}=3.4872*10^{-7},(Y_{\nu})_{33}=1.7208*10^{-6},
\nonumber\\&&L_4={3\over2},~\lambda_{N^c}=1.
\end{eqnarray}

To simplify the discussion of the numerical result, we assume the following relations
\begin{eqnarray}
&&(A_l)_{ii}=AL,(A'_l)_{ii}=A'_L, (A_{N^c})_{ii}=(A_N)_{ii}=AN,
(m_{\tilde{L}}^2)_{ii}=(m_{\tilde{R}}^2)_{ii}=S_m^2, \nonumber\\&&(m_{\tilde{N}^c}^2)_{ii}=M_{sn}^2,
(m_{\tilde{L}}^2)_{ij}=(m_{\tilde{R}}^2)_{ij}=M_{L_f},i\neq j,(i,j=1,2,3)
\end{eqnarray}
We choose the parameters $AL=-2$TeV, $A'_L=300$GeV, $M_{sn}=1$TeV, $\tan{\beta_L}=\bar{v}_L / v_L$ and $ V_{L_t}=\sqrt{\bar{v}_L^2 + v_L^2}$. $m_1$ represents the mass of gaugino in $U(1)$ and $m_2$ represents the mass of gaugino in $SU(2)$. Generally, the non-diagonal elements of the parameters are defined as zero unless we specially emphasize.
\subsection{$Z\rightarrow e\mu$}
The experimental upper bound for the branching ratio of $Z\rightarrow e\mu$ is around $7.5\times10^{-7}$. The parameter $m_1$ is related with the mass matrix of neutralino, which means the contributions from neutralino-slepton can be influenced by the parameter $m_1$. With $g_L=0.3$, $S_m=1$TeV, $AN=-500$GeV, $m_2=1$TeV, $M_{L_f}=1\times10^{5}$ GeV$^2$ and $\tan{\beta}=15$, we plot the results versus $m_1$  in Fig.2. As $m_1>0$, the results decrease with increasing $m_1$. However, the results are in the region $(3.0\times10^{-9}\sim3.5\times10^{-9})$ and the effect of $m_1$ is small.

As a more sensitive parameter, $m_2$ not only presents in the mass matrix of neutralino, but also in the mass matrix of chargino. This parameter affects the numerical results through the neutralino-slepton and chargino-sneutrino contributions. In Fig.3, we show the effects from $m_2$ with $g_L=0.2$, $S_m=1$TeV, $AN=-500$GeV, $\tan{\beta}=15$ and $M_{L_f}=1\times10^{5}$ GeV$^2$. And we plot the solid line, dotted line and dashed line respectively with $m_1=500 (1000, 2000)$GeV. These three lines all become small quickly with the increasing $m_2$. It implies that $m_2$ is a relatively sensitive parameter to the numerical results.
\begin{figure}[ht]
\includegraphics[width=10cm]{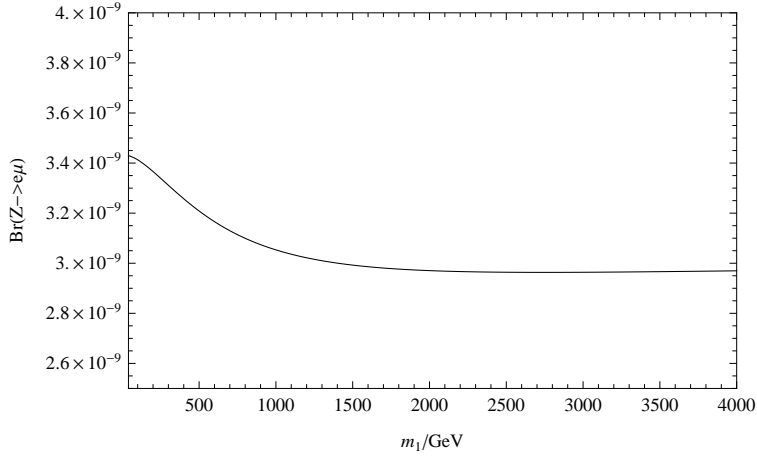}
\caption{\label{fig2}With $g_L=0.3,\; S_m=1$TeV, $AN=-500$GeV, $m_2=1$TeV, $M_{L_f}=1\times10^{5}$ GeV$^2$ and $\tan{\beta}=15$, the contributions to  $Br(Z\rightarrow e\mu)$ versus $m_1$ are plotted by the solid line.}
\end{figure}
\begin{figure}[ht]
\includegraphics[width=10cm]{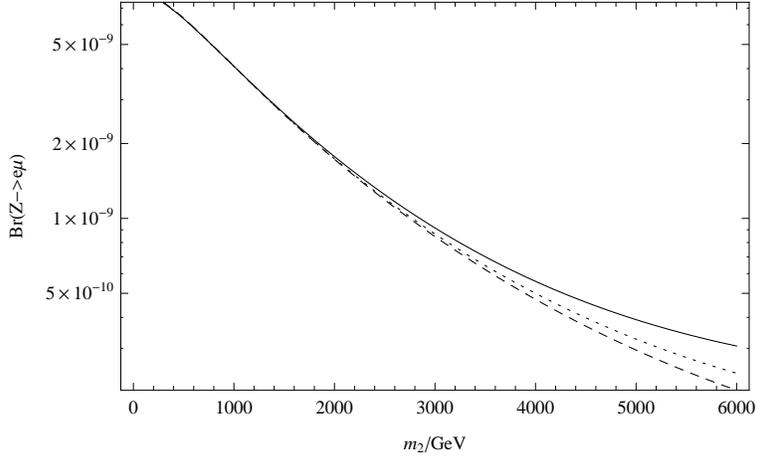}
\caption{\label{fig3}With $g_L=0.2,\; S_m=1$TeV, $AN=-500$GeV, $\tan{\beta}=15$, $M_{L_f}=1\times10^{5}$ GeV$^2$ and $m_1=500(1000,1500)$GeV, $Br(Z\rightarrow e\mu)$ versus $m_2$ are plotted by the solid line, dotted line and dashed line respectively.}
\end{figure}

The parameters $g_L$, $\tan{\beta}_L$ and $V_{L_t}$ all present in the mass squared matrices of sleptons, sneutrinos and lepton neutralinos. Therefore, these three parameters affect the results through slepton-neutrino, sneutrinos-chargino and slepton-lepton neutralino contributions. We choose the parameters $m_1=500$GeV, $m_2=1$TeV, $S_m=1$TeV, $AN=500$GeV and $\tan{\beta=15}$. As $V_{L_t}=3$TeV, we plot the allowed results with $\tan{\beta}_L$ versus $g_L$ in Fig.4. Obviously, when the value of $g_L$ is large enough, the value of $\tan{\beta}_L$ approaches 1.
When $g_L\leq0.3$, the parameter $\tan{\beta}_L$ can vary in the region of 0$\sim$2. It implies that $g_L$ is a sensitive parameter to the numerical results.
As $\tan{\beta}_L=2$, $g_L$ versus $V_{L_t}$ are scanned in Fig.5. We find that the allowed scope of $V_{L_t}$ shrinks and the value of $V_{L_t}$ decreases with the enlarging $g_L$. Therefore, the value of $g_L$ should not be large. Generally, we take $0.05\leq g_L\leq0.3$ and $V_{L_t}\sim3$TeV in our numerical calculations.
\begin{figure}[ht]
\includegraphics[width=10cm]{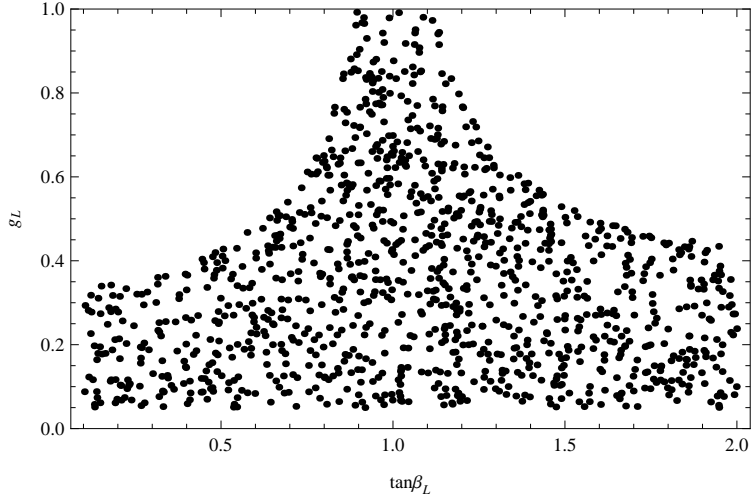}
\caption{\label{fig4}With $m_1=500$GeV, $m_2=1$TeV, $S_m=1$TeV, $AN=500$GeV, $\tan{\beta=15}$ and $V_{L_t}=3$TeV, the allowed parameter space in the plane of $\tan{\beta}_L$ versus $g_L$ for $Br(Z\rightarrow e\mu)$.}
\end{figure}
\begin{figure}[ht]
\includegraphics[width=10cm]{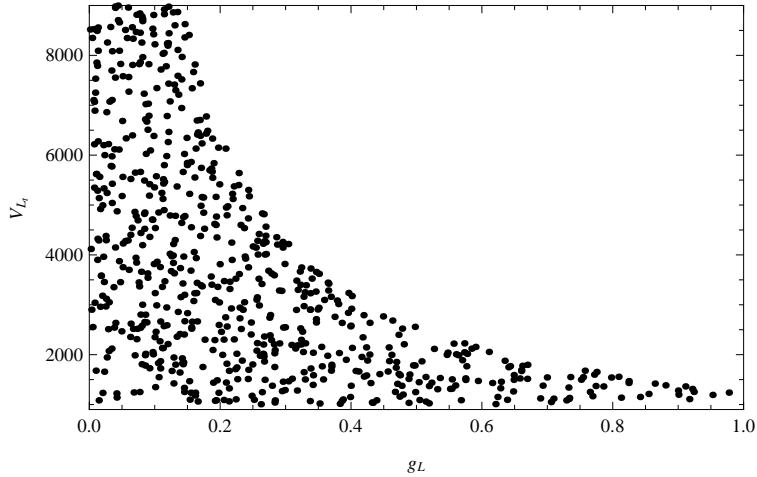}
\caption{\label{fig5}For $Br(Z\rightarrow e\mu)$, the allowed parameter space in the plane of $g_L$ versus $V_{L_t}$ with $m_1=500$GeV, $m_2=1$TeV, $S_m=1$TeV, $AN=500$GeV, $\tan{\beta=15}$ and $\tan{\beta}_L=2$.}
\end{figure}
\subsection{$Z\rightarrow e\tau$}
In the similar way, the CLFV process $Z\rightarrow e\tau$ is numerically studied and its experimental upper bound is around $9.8\times10^{-6}$. As discussed in the previous part, $g_L$ can affect the contribution strongly through the masses of sleptons, sneutrinos and lepton neutralinos. $S_m$ is the diagonal element of $m_{\tilde{L}}^2$ and $m_{\tilde{R}}^2$ in the slepton mass matrix, which can affect slepton-neutralino and slepton-lepton neutralino contributions in the CLFV process. Using the parameters $m_1=500$GeV, $m_2=1$TeV, $AN=-500$GeV, $\tan{\beta}=12$ and $M_{L_f}=1\times10^{5}$ GeV$^2$, we study the branching ratio versus $S_m$ with $g_L=0.1(0.15, 0.2)$ in Fig.6, and the results are plotted by the solid line, dotted line and dashed line respectively. These three lines decrease quickly with $S_m$ enlarging from 1000GeV to 2500GeV, which indicates that $S_m$ is a very sensitive parameter to the numerical results. When $S_m > 2500$GeV, the results decrease slowly and the branching ratios are around $(10^{-9}\sim10^{-10})$.
\begin{figure}[ht]
\includegraphics[width=10cm]{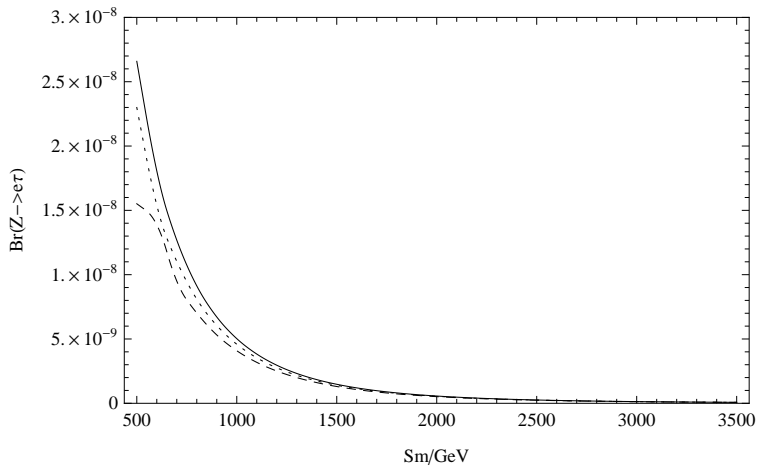}
\caption{\label{fig6}With $m_1=500$GeV, $m_2=1$TeV, $AN=-500$GeV, $\tan{\beta}=12$, $M_{L_f}=1\times10^{5}$ GeV$^2$ and $g_L=0.1(0.15, 0.2)$, the contributions to  $Br(Z\rightarrow e\tau)$ versus $S_m$ are plotted by the solid line, dotted line and dashed line respectively.}
\end{figure}

Then we study the process with the parameters $M_{L_f}$ and $m_2$. As $S_m=\sqrt{2}$TeV, $g_L=0.2$, $m_1=500$GeV, $AN=500$GeV, $\tan{\beta}=12$, we study the results versus $M_{L_f}$ with $m_2=1 (1.5, 2)$ TeV in Fig.7, and the results are plotted by the solid line, dotted line and dashed line respectively. As $M_{L_f}=0$, the branching ratio for $Z\rightarrow e\tau$ is almost zero, but the results increase sharply when  $|M_{L_f}| > 0$. We can know that non-zero $M_{L_f}$ is a sensitive parameter and has strong affection on the lepton flavor violation.
\begin{figure}[ht]
\includegraphics[width=10cm]{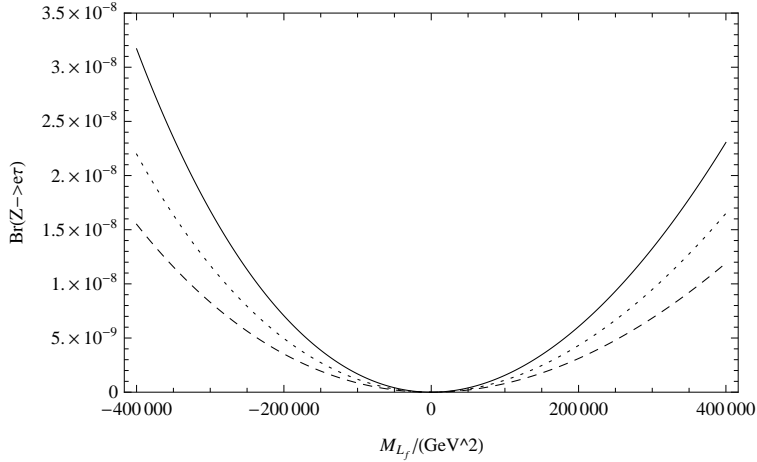}
\caption{\label{fig7}With $S_m=\sqrt{2}$TeV, $g_L=0.2$, $m_1=500$GeV, $AN=500$GeV, $\tan{\beta}=12$ and $m_2=1 (1.5, 2)$TeV, the contributions to  $Br(Z\rightarrow e\tau)$ versus $M_{L_f}$ are plotted by the solid line, dotted line and dashed line respectively.}
\end{figure}
\subsection{$Z\rightarrow \mu\tau$}
The experimental upper bound for the CLFV process $Z\rightarrow \mu\tau$ is $1.2\times10^{-5}$, which is about one order larger than the process $Z\rightarrow e\mu$. The parameter $AN$  presents in the sneutrino mass matrix and affects sneutrino-chargino contributions. Supposing $m_1=500$GeV, $m_2=1$TeV, $g_L=0.2$, $S_m=1$TeV, $M_{L_f}=1\times10^{5}$ GeV$^2$ and $\tan{\beta}=1(2,3)$, we plot the results with the $AN$ in Fig.8. As $AN\leq4$TeV, the branching ratios are around $4\times10^{-9}$; As $AN>4$TeV, these three lines increase quickly and $AN$ has an obvious influence on the numerical results.

After that, the effects from parameter $\tan{\beta}$ are studied.
$\tan{\beta}$ is related with $v_u$ and $v_d$, and almost appears in all mass matrices of CLFV processes. With $m_1=500$GeV, $m_2=1$TeV, $S_m=1$TeV, $g_L=0.3$, $M_{L_f}=-1\times10^{5}$ GeV$^2$ and $AN=500$GeV, Fig.9 is plotted to show the results along with the parameter $\tan{\beta}$. It indicates that the results do not change significantly. In the range of $\tan{\beta}=(0\sim3)$, we find that the branching ratio decreases slightly; As $\tan{\beta}>3$, the result is stable and around $3.7\times10^{-9}$.
\begin{figure}[ht]
\includegraphics[width=10cm]{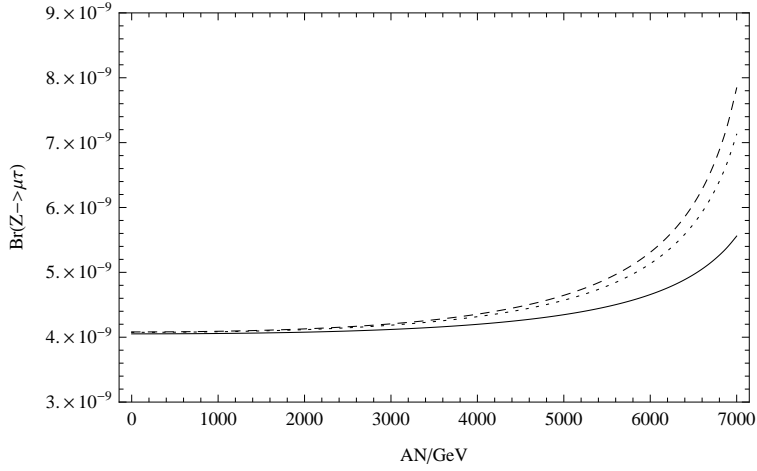}
\caption{\label{fig8}With $m_1=500$GeV, $m_2=1$TeV, $g_L=0.2,\; S_m=1$TeV, $M_{L_f}=1\times10^{5}$ GeV$^2$ and $ \tan{\beta}=1(2,3)$, the contributions to  $Br(Z\rightarrow \mu\tau)$ versus $AN$ are plotted by the solid line, dotted line and dashed line respectively.}
\end{figure}
\begin{figure}[ht]
\includegraphics[width=10cm]{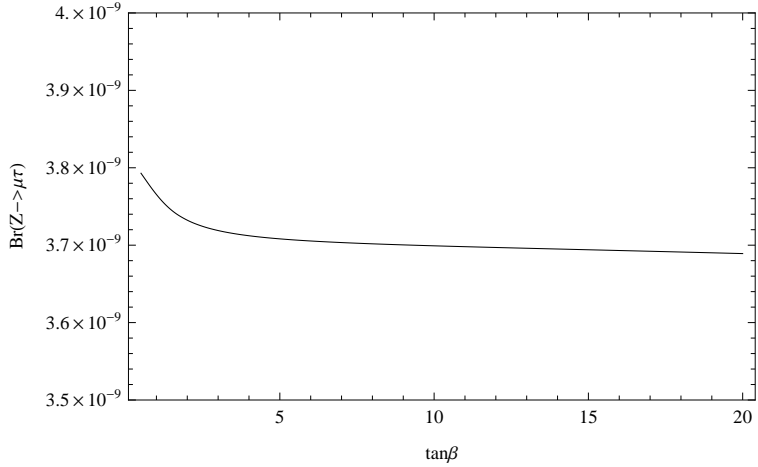}
\caption{\label{fig9}For $Br(Z\rightarrow \mu\tau)$, with $m_1=500$GeV, $m_2=1$TeV, $g_L=0.3,\; S_m=1$TeV, $M_{L_f}=-1\times10^{5}$ GeV$^2$, $AN=500$GeV, the results versus $\tan{\beta}$ are plotted by the solid line.}
\end{figure}
\section{discussion and conclusion\label{sec5}}
In this paper, we study the CLFV processes $Z\rightarrow l_i^{\pm}l_j^{\mp}$ in the BLMSSM. Compared with the MSSM with R-party conservation, there are new parameters and new contributions to the CLFV processes in the BLMSSM. For instance: 1. Three heavy neutrinos are introduced in this model. However, the new contributions from these particles are tiny, because the couplings of these particles are suppressed by tiny neutrino Yukawa $Y_\nu$.
2. Three new scalar neutrinos are introduced in this model. Considering the mass squared matrix of the sneutrinos in Eq.(10),
we find that the contributions from ${\cal M}^2_{\tilde{n}}(\tilde{\nu}_I\tilde{N}_J^c)$ can be neglect due to the tiny Yakawa couplings $Y_{\nu}$.  The effects from ${\cal M}^2_{\tilde{n}}(\tilde{\nu}_{I}^*\tilde{\nu}_{J})$ and ${\cal M}^2_{\tilde{n}}(\tilde{N}_I^{c*}\tilde{N}_J^c)$ play very important roles.
Although the diagonal elements of $(m^2_{\tilde{L}})_{IJ}$ and $(m^2_{\tilde{N}^c})_{IJ}$ suppress the contributions, the non-diagonal element $M_{L_f}$ of $(m^2_{\tilde{L}})_{IJ}$ leads to strong mixing for sneutrinos of different generations.
Therefore, the nonzero $M_{L_f}$ enhances lepton flavor violation and leads to large results.
3. Lepton neutralinos $\chi_L^0$ are the new particles introduced in our work.
The numerical results can be influenced by the slepton-lepton neutralinos contributions. As the non-diagonal elements, $(\mathcal{M}^2_L)_{LR}$ are not small and can obviously improve the lepton flavor violation effects.
Furthermore, the parameters $(m^2_{\tilde{L}})_{IJ}$ and $(m^2_{\tilde{R}})_{IJ}$
respectively exist in $(\mathcal{M}^2_L)_{LL}$ and $(\mathcal{M}^2_L)_{RR}$.
It indicates that the non-diagonal element $M_{L_f}$ of $(m^2_{\tilde{L}})_{IJ}$ and $(m^2_{\tilde{R}})_{IJ}$
leads to strong mixing for sleptons. Therefore, $(\mathcal{M}^2_L)_{LR}$ and $M_{L_f}$ influence our results strongly.

In our used parameter space, the numerical results show that the rates for $Br(Z\rightarrow l_i^{\pm}l_j^{\mp})$ can almost reach the present experimental upper bounds. The numerical analyses indicate that parameters $m_1,\; m_2,\; g_L,\; M_{L_f},\; S_m,\; AN$ and $\tan{\beta}$ are important. The sensitive parameters are $g_L$, $M_{L_f}$ and $S_m$ and they affect the results obviously. We hope the experiment results for $Z\rightarrow l_i^{\pm}l_j^{\mp}$ can be detected in the near future.

\begin{acknowledgments}
Supported by the Major Project of
NNSFC (No. 11535002, No. 11605037, No. 11647120, No. 11275036),
the Natural Science Foundation of Hebei province with Grant
No. A2016201010 and No. A2016201069, and the Natural Science Fund of
Hebei University with Grants No. 2011JQ05 and No. 2012-
242, Hebei Key Lab of Optic-Electronic Information and
Materials, the midwest universities comprehensive strength
promotion project.
\end{acknowledgments}

\appendix

\section{The coupling coefficients\label{app-coupling}}
The concrete forms of coupling coefficients corresponding to Fig.1(1)$\sim$Fig.1(9) are shown as:

Fig.1(1): $S_1=\tilde{\nu}_n,S_2=\tilde{\nu}_m,F=\chi^c$
\begin{eqnarray}
&&H_L^{S_2F\bar{l}_i}(1)=-Y_l^{Im*}Z_-^{2k}Z_{\tilde{\nu}}^{Im},
\nonumber\\&&H_R^{S_2F\bar{l}_i}(1)=-[\frac{e}{s_w}Z_+^{1k*}Z_{\tilde{\nu}}^{Im}+Y_{\nu}^{Im*}Z_+^{2k*}Z_{\tilde{\nu}}^{(I+3)m}],
\nonumber\\&&H_L^{S_1^*l_j\bar{F}}(1)=-[\frac{e}{s_w}Z_+^{1k}Z_{\tilde{\nu}}^{Jn*}+Y_{\nu}^{Jn}Z_+^{2k}Z_{\tilde{\nu}}^{(J+3)n*}],
\nonumber\\&&H_R^{S_1^*l_j\bar{F}}(1)=-{Y_l^{Jn}Z_-^{2k*}Z_{\tilde{\nu}}^{Jn*}},
\nonumber\\&&H^{ZS_1S_2^*}(1)=\frac{e}{2s_wc_w}Z_{\tilde{\nu}}^{Km*}Z_{\tilde{\nu}}^{Kn}.
\end{eqnarray}

Fig.1(2): $S_1=\tilde{L}_n,S_2=\tilde{L}_m,F=\chi^0$
\begin{eqnarray}
&&H_L^{S_2F\bar{l}_i}(2)=\frac{-\sqrt{2}e}{c_w}Z_L^{(I+3)m*}Z_N^{1k}+Y_l^{I*}Z_L^{Im*}Z_N^{3k},
\nonumber\\&&H_R^{S_2F\bar{l}_i}(2)=\frac{e}{\sqrt{2}s_wc_w}Z_L^{Im*}(Z_N^{1k*}s_w+Z_N^{2k*}c_w)
+Y_l^{I*}Z_L^{(I+3)m*}Z_N^{3k*},
\nonumber\\&&H_L^{S_1^*l_j\bar{F}}(2)=\frac{e}{\sqrt{2}s_wc_w}Z_L^{Jn}(Z_N^{1k}s_w+Z_N^{2k}c_w)
+Y_l^JZ_L^{(J+3)n}Z_N^{3k},
\nonumber\\&&H_R^{S_1^*l_j\bar{F}}(2)=\frac{-\sqrt{2}e}{c_w}Z_L^{(J+3)n}Z_N^{1k*}
+Y_l^JZ_L^{Jn}Z_N^{3k*},
\nonumber\\&&H^{ZS_1S_2^*}(2)=-\frac{e}{2s_wc_w}(Z_L^{Km}Z_L^{Kn*}-2s_w^2\delta^{mn}).
\end{eqnarray}

Fig.1(3): $S_1=\tilde{L}_n,S_2=\tilde{L}_m,F=\chi_L^0$
\begin{eqnarray}
&&H_L^{S_2F\bar{l}_i}(3)=-\sqrt{2}g_LZ_{N_L}^{1k}Z_L^{(I+3)m*},
\nonumber\\&&H_R^{S_2F\bar{l}_i}(3)=\sqrt{2}g_LZ_{N_L}^{1k*}Z_L^{Im*},
\nonumber\\&&H_L^{S_1^*l_j\bar{F}}(3)=\sqrt{2}g_LZ_{N_L}^{1k}Z_L^{Jn},
\nonumber\\&&H_R^{S_1^*l_j\bar{F}}(3)=-\sqrt{2}g_LZ_{N_L}^{1k*}Z_L^{(J+3)n},
\nonumber\\&&H^{ZS_1S_2^*}(3)=H^{ZS_1S_2^*}(2).
\end{eqnarray}

Fig.1(4): $S_1=H^{\pm}(G^{\pm}),S_2=H^{\pm}(G^{\pm}),F=\nu$
\begin{eqnarray}
&&H_L^{S_2F\bar{l}_i}(4,H)=-\sin{\beta}Y_l^{Ik}U_{\nu}^{Ik},
\nonumber\\&&H_R^{S_2F\bar{l}_i}(4,H)=-\cos{\beta}Y_{\nu}^{Ik*}U_{\nu}^{(I+3)k},
\nonumber\\&&H_L^{S_1^*l_j\bar{F}}(4,H)=-\cos{\beta}Y_{\nu}^{Jk}U_{\nu}^{(J+3)k*},
\nonumber\\&&H_R^{S_1^*l_j\bar{F}}(4,H)=-\sin{\beta}Y_l^{Jk*}U_{\nu}^{Jk*},
\nonumber\\&&H^{ZS_1S_2^*}(4,H)=-e\delta^{mn}\frac{c_w^2-s_w^2}{2s_wc_w},
\nonumber\\&&H_L^{S_2F\bar{l}_i}(4,G)=\cos{\beta}Y_l^{Ik}U_{\nu}^{Ik},
\nonumber\\&&H_R^{S_2F\bar{l}_i}(4,G)=-\sin{\beta}Y_{\nu}^{IK*}U_{\nu}^{(I+3)k},
\nonumber\\&&H_L^{S_1^*l_j\bar{F}}(4,G)=-\sin{\beta}Y_{\nu}^{Jk}U_{\nu}^{(J+3)k*},
\nonumber\\&&H_R^{S_1^*l_j\bar{F}}(4,G)=\cos{\beta}Y_l^{Jk*}U_{\nu}^{Jk*},
\nonumber\\&&H^{ZS_1S_2^*}(4,G)=H^{ZS_1S_2^*}(4,H).
\end{eqnarray}

Fig.1(5): $F_1=\chi^c_n,F_2=\chi^c_m,S=\tilde{\nu}$
\begin{eqnarray}
&&H_L^{SF_2\bar{l}_i}(5)=-Y_l^{Ik*}Z_-^{2m}Z_{\tilde{\nu}}^{Ik},
\nonumber\\&&H_R^{SF_2\bar{l}_i}(5)=-[\frac{e}{s_w}Z_+^{1m*}Z_{\tilde{\nu}}^{Ik}+Y_{\nu}^{Ik*}Z_+^{2m*}Z_{\tilde{\nu}}^{(I+3)k}],
\nonumber\\&&H_L^{ZF_1\bar{F}_2}(5)=-\frac{e}{2s_wc_w}[Z_+^{1m*}Z_+^{1n}+\delta^{mn}(c_w^2-s_w^2)],
\nonumber\\&&H_R^{ZF_1\bar{F}_2}(5)=-\frac{e}{2s_wc_w}[Z_-^{1m}Z_-^{1n*}+\delta^{mn}(c_w^2-s_w^2)],
\nonumber\\&&H_L^{S^*l_j\bar{F}_1}(5)=-[\frac{e}{s_w}Z_+^{1n}Z_{\tilde{\nu}}^{Jk*}+Y_{\nu}^{Jk}Z_+^{2n}Z_{\tilde{\nu}}^{(J+3)k*}],
\nonumber\\&&H_R^{S^*l_j\bar{F}_1}(5)=-{Y_l^{Jk}Z_-^{2n*}Z_{\tilde{\nu}}^{Jk*}}.
\end{eqnarray}

Fig.1(6): $F_1=\chi^0_n,F_2=\chi^0_m,S=\tilde{L}$
\begin{eqnarray}
&&H_L^{SF_2\bar{l}_i}(6)=\frac{-\sqrt{2}e}{c_w}Z_L^{(I+3)k*}Z_N^{1m}+Y_l^{I*}Z_L^{Ik*}Z_N^{3m},
\nonumber\\&&H_R^{SF_2\bar{l}_i}(6)=\frac{e}{\sqrt{2}s_wc_w}Z_L^{Ik*}(Z_N^{1m*}s_w+Z_N^{2m*}c_w)
+Y_l^{I*}Z_L^{(I+3)k*}Z_N^{3m*},
\nonumber\\&&H_L^{ZF_1\bar{F}_2}(6)=\frac{e}{2s_wc_w}(Z_N^{4m*}Z_N^{4n}-Z_N^{3m*}Z_N^{3n}),
\nonumber\\&&H_R^{ZF_1\bar{F}_2}(6)=-\frac{e}{2s_wc_w}(Z_N^{4m}Z_N^{4n*}-Z_N^{3m}Z_N^{3n*}),
\nonumber\\&&H_L^{S^*l_j\bar{F}_1}(6)=\frac{e}{\sqrt{2}s_wc_w}Z_L^{Jk}(Z_N^{1n}s_w+Z_N^{2n}c_w)
+Y_l^JZ_L^{(J+3)k}Z_N^{3n},
\nonumber\\&&H_R^{S^*l_j\bar{F}_1}(6)=\frac{-\sqrt{2}e}{c_w}Z_L^{(J+3)k}Z_N^{1n*}
+Y_l^JZ_L^{Jk}Z_N^{3n*}.
\end{eqnarray}

Fig.1(7): $F_1=\nu_n,F_2=\nu_m,S=H^{\pm}(G^{\pm})$
\begin{eqnarray}
&&H_L^{SF_2\bar{l}_i}(7,H)=-\sin{\beta}Y_l^{Im}U_{\nu}^{Im},
\nonumber\\&&H_R^{SF_2\bar{l}_i}(7,H)=-\cos{\beta}Y_{\nu}^{Im*}U_{\nu}^{(I+3)m},
\nonumber\\&&H_L^{ZF_1\bar{F}_2}(7,H)=-\frac{e}{2s_wc_w}U_{\nu}^{Km*}U_{\nu}^{Kn},
\nonumber\\&&H_R^{ZF_1\bar{F}_2}(7,H)=0,
\nonumber\\&&H_L^{S^*l_j\bar{F}_1}(7,H)=-\cos{\beta}Y_{\nu}^{Jn}U_{\nu}^{(J+3)n*},
\nonumber\\&&H_R^{S^*l_j\bar{F}_1}(7,H)=-\sin{\beta}Y_l^{Jn*}U_{\nu}^{Jn*},
\nonumber\\&&H_L^{SF_2\bar{l}_i}(7,G)=\cos{\beta}Y_l^{Im}U_{\nu}^{Im},
\nonumber\\&&H_R^{SF_2\bar{l}_i}(7,G)A_R=-\sin{\beta}Y_{\nu}^{Im*}U_{\nu}^{(I+3)m},
\nonumber\\&&H_L^{ZF_1\bar{F}_2}(7,G)=H_L^{ZF_1\bar{F}_2}(7,H),
\nonumber\\&&H_R^{ZF_1\bar{F}_2}(7,G)=0,
\nonumber\\&&H_L^{S^*l_j\bar{F}_1}(7,G)=-\sin{\beta}Y_{\nu}^{Jn}U_{\nu}^{(J+3)n*},
\nonumber\\&&H_R^{S^*l_j\bar{F}_1}(7,G)=\cos{\beta}Y_l^{Jn*}U_{\nu}^{Jn*}.
\end{eqnarray}

Fig.1(8): $W_1=W_1,W_2=W_2,F=\nu$
\begin{eqnarray}
&&H_L^{W_2F\bar{l}_i}(8)=-\frac{e}{\sqrt{2}s_w}U_{\nu}^{Ik},
\nonumber\\&&H_L^{W_1^*l_j\bar{F}}(8)=-\frac{e}{\sqrt{2}s_w}U_{\nu}^{Jk*},
\nonumber\\&&H^{ZW_1W_2^*}(8)=\frac{ec_w}{s_w},
\nonumber\\&&H_R^{W_2F\bar{l}_i}(8)=H_R^{W_1^*l_j\bar{F}}(8)=0.
\end{eqnarray}

Fig.1(9): $F_1=\nu_n,F_2=\nu_m,W=W$
\begin{eqnarray}
&&H_L^{WF_2\bar{l}_i}(9)=-\frac{e}{\sqrt{2}s_w}U_{\nu}^{Im},
\nonumber\\&&H_L^{ZF_1\bar{F}_2}(9)=-\frac{e}{2s_wc_w}U_{\nu}^{Km*}U_{\nu}^{Kn},
\nonumber\\&&H_L^{\bar{F}_1l_jW^*}(9)=-\frac{e}{\sqrt{2}s_w}U_{\nu}^{Jn*},
\nonumber\\&&H_R^{WF_2\bar{l}_i}(9)=H_R^{ZF_1\bar{F}_2}(9)=H_R^{\bar{F}_1l_jW^*}(9)=0.
\end{eqnarray}

\end{document}